\documentclass[aps,pre,notitlepage]{revtex4-1}
\usepackage{amsmath,amssymb,bm}
\usepackage{graphicx}
\usepackage{subfigure}
\usepackage{tkz-berge}
\usepackage{graphicx}

\usepackage{kantlipsum}
\usepackage{graphicx}
\usepackage{subfigure}
\usepackage{cancel}

\begin{document}
\title{
Absorbing transition in a coevolution model with node and link states in an adaptive network: Network fragmentation transition at criticality}

\author{Meghdad Saeedian$^{1,2}$, Maxi San Miguel$^3$, Raul Toral$^3$}

\affiliation{$^1$\textit{Dipartimento di Fisica ``G. Galilei'', Universit\`{a} di Padova, Via Marzolo 8, 35131 Padova, Italy}}
\affiliation{$^2$\textit{School of Physics, Institute for Research in Fundamental Sciences (IPM), Tehran, 19395-5531, Iran}}
\affiliation{$^3$\textit{IFISC Instituto de Fisica Interdisciplinar y Sistemas Complejos (CSIC-UIB), Campus Universitat Illes Balears, E-07122 Palma de Mallorca, Spain}}

\date{\today}
\begin{abstract}

We consider a general model in which there is a coupled dynamics of node states and links states in a network. This coupled dynamics coevolves with dynamical changes of the topology of the network caused by a link rewiring mechanism. Such coevolution model features the interaction of the local dynamics of node and link states with the nonlocal dynamics of link-rewiring in a random network. The coupled dynamics of the states of the nodes and the links produces by itself an absorbing phase transition which is shown to be robust against the link rewiring mechanism. However, the dynamics of the network gives rise to significant physical changes, specially in
the limit in which some links do not change state but are always rewired: First a network fragmentation occurs at the critical line of the absorbing transition, and only at this line, so that fragmentation is a manifestation of criticality. Secondly, in the active phase of the absorbing transition, finite-size fluctuations take the system to a single network component consensus phase, while other configurations are possible in the absence of rewiring. In addition, this phase is reached after a survival time that scales linearly with system size, while the survival time scales exponentially with system size when there is no rewiring. A social interpretation of our results contribute to the description of processes of emergence of social fragmentation and polarization.

\end{abstract}

\maketitle

\section*{Introduction}
Modeling a complex system of interacting agents requires specifying the network of interactions and the state of the agents, represented as nodes of the network. The links of the network can also have a state, representing for instance attractive or repulsive interactions. In addition, the network might not be fixed, but adaptive with a time dependent topology. A general dynamical model includes the coupled dynamics of the states of the nodes, the states of the links and the topology of the network. In terms of social collective emergent properties a key question is if the asymptotic state of the dynamics describes some sort of consensus or alternatively, group formation with social fragmentation and/or polarization.

A first class of models are those that do not include link states, other than existence or non-existence of the link between two nodes. Among these, there are models that describe dynamics \emph{on} the network: evolution of the states of the nodes in a fixed network, while other describe dynamics \emph{of} the network, as for example dynamics of network formation with no reference to the state of the nodes. We use the term \emph{coevolution} models \cite{zimmermann2001cooperation,zimmermann2004coevolution} for those describing a coupled evolution of these two dynamics: node state dynamics and network topology dynamics, so that the structure of the network is no longer a given, but a variable. In these coevolution models the network plasticity is a parameter measuring the ratio of time scales of network evolution and evolution of the states of the nodes. In many cases coevolution models feature a network fragmentation transition \cite{vazquez2008generic,diakonova2014absorbing}, associated with an absorbing phase transition, in which a single component network fragments into several disconnected components. Coevolution models have been analyzed in the context of social differentiation \cite{eguiluz2005cooperation}, voter model and opinion formation \cite{vazquez2008generic,holme2006nonequilibrium,min2017fragmentation,min2019multilayer} or cultural polarization \cite{centola2007homophily}. They have also been used to describe empirical data on community structure of online games \cite{klimek2016dynamical} and more recently twitter data on echo chambers and polarization dynamics \cite{baumann2020modeling} as well as echo chambers on political parliamentary voting \cite{evans2018opinion}.

A different class of models are those that focus on the state of the link \cite{fernandez2012dynamics}. While the existence of a link in the context of opinion formation has been often interpreted as a positive interaction like friendship, trust or collaboration, it has also been argued that negative interactions are a major driving force in collective social behavior \cite{bliuc2015public,hutchings2019prejudice}. Therefore, considering positive and negative links should be an essential ingredient of interacting agents models. Links states were pioneered in Heider's social balance theory \cite{heider1946attitudes,heider2013psychology,antal2005dynamics,antal2006social,szell2010multirelational,marvel2011continuous}. Coevolution of link states and network topology, with no node state dynamics, has also been considered \cite{carro2014fragmentation}. More attention has been recently paid to the coupled dynamics of nodes and link states in a time independent network (fixed topology) in different contexts \cite{singh2014extreme,saeedian2017epidemic,carro2016coupled,pham2020effect,saeedian2019absorbing}. In particular, in \cite{saeedian2019absorbing}, inspired by processes of opinion formation, we considered an interacting agents model in which each node can be in either of two states and links can also be in either of two states associated with a friendly or unfriendly relation. Two nodes in a different state linked by a friendly relation or two nodes in the same state linked by an unfriendly relation are considered as unsatisfying pairs and they evolve to satisfying pairs by changing the state of one of the nodes or by changing the state of the link. This model exhibits an absorbing phase transition from a dynamically \textbf{active} state with persistent dynamics to a \textbf{frozen absorbing} configuration. The transition occurs for a critical value of the relative time scale for node and link state updates, that depends on the average degree of the network. The system also shows a finite-size topological transition associated with group splitting. Groups are here defined as a set of nodes in the same state connected by friendly links among themselves and by unfriendly links to the nodes in other groups. The existence of these groups, when they are poorly connected among them, reflects social fragmentation, with each group acting as an echo chamber. While \cite{saeedian2019absorbing} considers dyadic interactions, a related model in \cite{pham2020effect} implements an optimization dynamics of social balance triangular relations.

In this paper we go beyond the model of \cite{saeedian2019absorbing} by also considering adaptive dynamics of the network, so that we have a full dynamical model including dynamical evolution of the states of the nodes and the states of the links as well a time dependent network topology. This is a coevolution model, in the sense that the coupled dynamics of node-link states coevolves with dynamical changes of the topology of the network caused by a link rewiring mechanism. Note also that this model includes a coupling between the local mechanism of changes of node and link states and the nonlocal mechanism of random rewiring in the network.
We find in this model an absorbing phase transition which is rather independent of the network dynamics. However, network dynamics gives rise to significant new phenomena. In particular, link rewiring can produce a network fragmentation transition on the critical line of the absorbing transition and it modifies essentially the nature of the ac phase: a consensus phase can be reached by finite-size fluctuations.

The paper is organized as follows. First, we discuss our general coevolution model and its rate equation formulation. Next, we describe the absorbing transition predicted by the rate equations, as well as results on this transition obtained by Monte Carlo simulations. In the following sections we focus on the special case of largest rewiring probability, the approach to the absorbing states and calculation of survival times, and the topological transitions among the final frozen states. The final section summarizes our main results. An Appendix discusses the derivation of the rate equations.

\section*{Coevolution model: Coupling of node-link states local dynamics and non-local network dynamics.}
\begin{figure}[t]
\centering
\includegraphics[width=.8\textwidth]{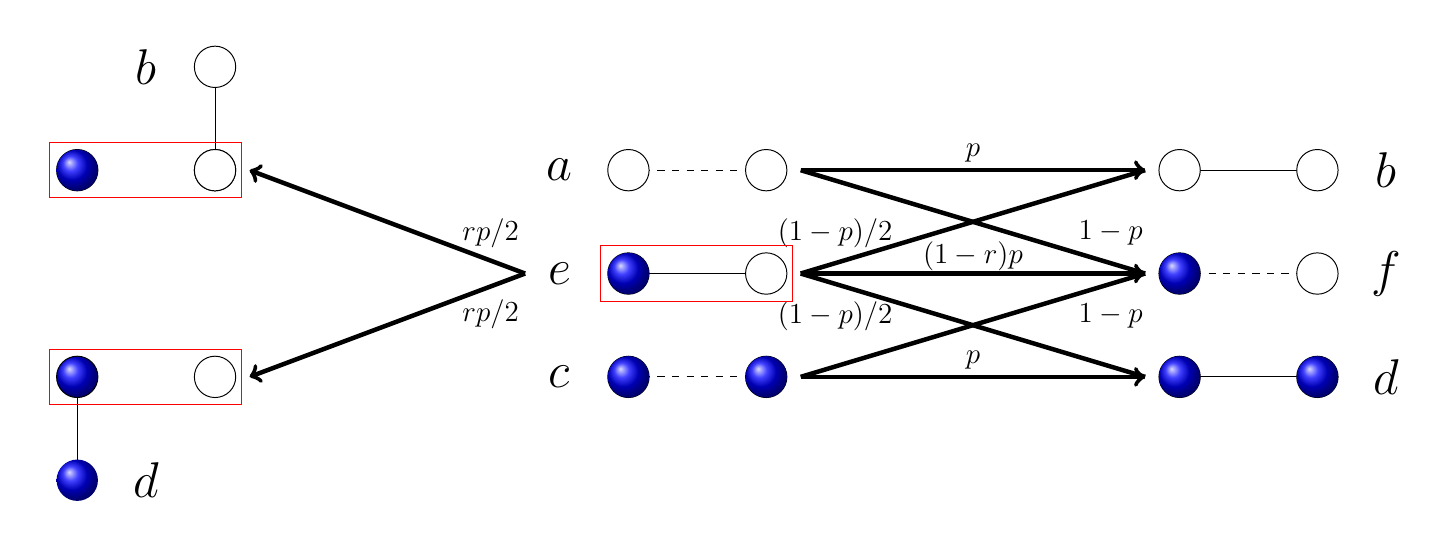}
\caption{All six possible configurations of pairs and the associated update events including update of states of nodes and links (rightward arrows) as local dynamics and rewiring the links (leftward arrows) as non-local dynamics. The pairs labeled by $a$, $c$ and $e$ are unsatisfying, while pairs $b$, $d$ and $f$ are satisfying. \textit{p} is the probability of update of state of links and $1-p$, the complementary probability, is the probability of node state update. \textit{r} is the probability of rewiring once the pair $e$ has been selected for link update.}
\label{update_rule_Co}
\end{figure}

We consider an initial uncorrelated random network with \textit{N} nodes and an average degree $\mu$ \cite{albert2002statistical}. Each node is endowed by a binary state value $1$ or $-1$. Each link also holds a binary state variable $+$ or $-$ representing, respectively, attractive or repulsive interactions. In an opinion formation model, the state of the nodes represent two different opinions and attractive or repulsive interactions stand for friendly or unfriendly relations. The states of the nodes are represented in the figures by filled-in blue and white circles, while the states of the links are represented by solid ($+$) or dashed ($-$) lines. We consider repulsive links connecting nodes with different (resp. the same) state and attractive links connecting nodes with the same (resp. different) state as \textbf{\textit{satisfying}} (resp. \textbf{\textit{unsatisfying}}) relations. In Fig.~\ref{update_rule_Co}, we show all $6$ possible types of pairs and identify \textit{a}, \textit{e} and \textit{c} as unsatisfying pairs and \textit{b}, \textit{d} and \textit{f} as satisfying ones \cite{saeedian2019absorbing}.

We implement a dynamical model that tends to minimize the number of unsatisfying pairs. The change from unsatisfying pairs to satisfying ones is done through two different mechanisms:

\textbf{(i)}\textbf{\textit{Flipping node or link states}} (rightward arrows in Fig.~\ref{update_rule_Co}): This is a local mechanism in which an unsatisfying pair is turned into a satisfying one by changing the state of one of the nodes or the state of the link. With probability \textit{p} a link state is changed, and with the complimentary probability (1-\textit{p}) a node state is changed. This gives rise to a coupled dynamics of node and link states in which $p$ measures the ratio of time scales of node and link evolution. The pair \textit{a} is turned into a satisfying pair \textit{b} via a link update with probability \textit{p} or it is turned into a satisfying pair \textit{f} through a node update with complimentary probability (1-\textit{p}), as shown in Fig.~\ref{update_rule_Co}. Note that in the node update \textit{a} to \textit{f}, it doesn't matter which of the nodes in the pair \textit{a} changes its state, both result in \textit{f}. The pair \textit{c} is similarly transformed into a satisfying pair: a link update transforms it into the pair \textit{d} and a node update (from any side) transforms it into the pair \textit{f}. For the case of the unsatisfying pair \textit{e}, a link update with probability \textit{p} transforms it into the satisfying pair \textit{f}. With complimentary probability (\textit{1-p}), a node update takes place on pair \textit{e}; either its white node with probability $\frac{1-p}{2}$ flips its state and \textit{e} becomes \textit{d}, or alternatively the blue node with probability $\frac{1-p}{2}$ changes its state and \textit{e} becomes \textit{b}.

\textbf{(ii)}\textbf{\textit{Link rewiring}} (leftward arrows in Fig.~\ref{update_rule_Co}). In this nonlocal mechanism \cite{vazquez2008generic}, one of the two nodes, chosen at random, of an unsatisfying pair of type \textit{e} breaks the existing link and reconnects to another node of its own type, again chosen randomly amongst all possible nodes. Hence the unsatisfying pair of type \textit{e} can either become a satisfying pair of type \textit{b} or of type \textit{d} according to which one of the two nodes (white or blue, respectively) is selected to find a matching node in the network. Summing up, when an $e$-type pair has been chosen, the following actions can occur: (i) change the state of the link from attractive to repulsive, with probability $(1-r)p$; (ii) the node holding the blue opinion turns into white, with probability $\frac{1-p}{2}$; (iii) the node holding the white opinion turns into blue, with probability $\frac{1-p}{2}$; (iv) the node holding the white opinion breaks the link and binds to another white node in the network, with probability $\frac{rp}{2}$, (v) the node holding the blue opinion breaks the link and binds to another blue node in the network, with probability $\frac{rp}{2}$.

The parameter \textit{r} sets the relative time scale of network evolution. In principle one could also think of a rewiring mechanism that does not change link state, but instead turns pairs \textit{a} and \textit{c} into a satisfying pair \textit{f}. We do not include this possibility here: In terms of an opinion formation model it corresponds to a search in the network to establish an unfriendly relation with an agent with an opposite opinion. It can be socially argued that this process would occur with a much smaller probability than the one considered here of searching to establish a friendly relation with an agent with the same opinion.

The three parameters of the model, in addition of system size $N$, are the probabilities $p$ and $r$ and the average degree of the network $\mu$. The limiting case $r=0$ was considered in \cite{saeedian2019absorbing}. In that case there is coupled evolution of node and links states by local dynamics, but there is no network dynamics. In our coevolving case, the network evolves by a nonlocal mechanism coupled to the evolution of node and link states. In the limiting case $p=0$, there is only dynamics of the states of the nodes and no network dynamics.
Still this limit does not correspond to the well known voter model in a fixed network \cite{vazquez2008analytical} where updates
from pairs \textit{a} and \textit{c} to pair \textit{f} do not occur. In the case $p=1$ there is only link dynamics
(including rewiring).
 
In a Monte Carlo implementation of this coevolution model, the system evolves asynchronously. A single pair is selected at random at each time step: If it is satisfying, nothing happens. If it is unsatisfying, it adapts according to the dynamics described in Fig.~\ref{update_rule_Co}. A Monte Carlo step is counted after a sequence of time steps equal to the total number of links in the network. In an Erd\H{o}s-R\'enyi network \cite{barrat2008dynamical} with $N$ nodes and average degree $\mu$ the total number of links is $L=\frac12 \mu N$, a quantity that is kept constant during the evolution. Defining $L_i$, \textit{$i\in$\{a, b, c, d, e, f\}} as the number of pairs of type $i$, the associated densities are $\rho_i=L_i/L$, satisfying the obvious normalization condition $\rho_{a}+\rho_{b}+\rho_{c}+\rho_{d}+\rho_{e}+\rho_{f}=1$.

The dynamical rules are such that the unsatisfying pairs ($\rho_{a}$, $\rho_{c}$ and $\rho_{e}$) are turned into satisfying. An absorbing configuration of the system \cite{hinrichsen2000non} is reached when there are no unsatisfying pairs, $\rho_{a}=\rho_{c}=\rho_{e}=0$. The system is then in a frozen state where there can be no further evolution. Note that, although at each step of the dynamics an unsatisfying pair becomes a satisfying pair, this does not mean that the total number of unsatisfying pairs decreases, because an update that converts a pair from unsatisfying to satisfying by changing an individual node state might also change the status of another pair, to which the node involved in the update also belongs to, from
satisfying to unsatisfying. The main question is when and how an absorbing or frozen configuration is reached. To answer this question we will analyze the dynamical evolution of the densities $\rho_i(t)$, \textit{$i\in$\{a, b, c, d, e, f\}}. This dynamics is analyzed by Monte Carlo simulations and by a set of rate equations.

\section*{Rate equations}
The densities $\{\rho_a,\rho_b,\rho_c,\rho_d,\rho_e,\rho_f\}$ obey six coupled differential rate equations (RE) describing the model sketched in Fig.~\ref{update_rule_Co}. These equations are derived in the \textit{Appendix}. Each equation indicates the time evolution of one of the densities. The rate equations are a combination of linear and nonlinear terms. The linear terms describe the change of densities due to the direct update of pairs shown in Fig.~\ref{update_rule_Co}, while the nonlinear terms are the outcome of the node update causing a change in the densities of the pairs linked to the updated node. There are two main assumptions in the derivation of these equations to be taken into account when comparing their predictions with Monte Carlo simulations. First, they are derived in the thermodynamic limit ($N\rightarrow\infty$), so that they cannot describe finite-size fluctuations. Second, they assume a random regular network in which all nodes have exactly $\mu$ random neighbors.

The rate equations have two sets of fixed points or stationary solutions. The first solution is given by
\begin{eqnarray} \label{first_set}
\rho_{a}^\text{st}&=&\rho_{c}^\text{st}=\frac{1-r}{2}\rho_{e}^\text{st}, \quad
\rho_{b}^\text{st}=\rho_{d}^\text{st}=\frac{\rho_{f}^\text{st}}{2(1-r)},\cr
\rho_{e}^\text{st}&=&\frac{-3+q(2-r)+2r}{(2-r)(2q(2-r)+r)},\cr
\rho_{f}^\text{st}&=&\frac{1-r}{2-r}-(1-r)\rho_{e}^\text{st},
\end{eqnarray}
where $q=(\mu-1)(1-p)$. This satisfies the normalization condition $\rho^\text{st}_a+\rho^\text{st}_b+\rho^\text{st}_c+\rho^\text{st}_d+\rho^\text{st}_e+\rho^\text{st}_f=1$. The second set of solutions is
\begin{eqnarray}
\rho_{a}^\text{st}= 0, \ \rho_{c}^\text{st}= 0, \ \rho_{e}^\text{st}=0, \quad (\rho^\text{st}_{b},\rho^\text{st}_{d},\rho^\text{st}_{f}) \rightarrow \textrm{arbitrary}, \label{second_set}
\end{eqnarray}
i.e. there are no unsatisfying pairs, and the relative fractions of satisfying pairs are arbitrary, but still need to fulfill the normalization condition $\rho^\text{st}_b+\rho^\text{st}_d+\rho^\text{st}_f=1$.

The first solution, Eq.~\eqref{first_set}, described by generally non-vanishing values of the densities that depend on on $r$, $p$ and $\mu$, is approached asymptotically on time and it is independent of initial conditions. Given that in this solution there exist unsatisfying pairs, it corresponds to a dynamically active state. The second set of solutions, Eq.~\eqref{second_set}, includes a continuous of solutions which depend on initial conditions. However, in all of them there are no unsatisfying pairs, and therefore they correspond to absorbing states with frozen dynamics.

\section*{Absorbing Transition}

The condition $\rho^\text{st}_e\ge0$ determines that the first solution, Eq.~\eqref{first_set}, only exists for $-3+(\mu-1)(1-p)(2-r)+2r\ge0$ or, for given $r$ and $\mu$, for $p\le p_{c}(\mu,r)$ with
\begin{eqnarray}\label{eq:pc}
p_{c}(\mu,r)=1-\frac{3-2r}{(2-r)(\mu-1)}.
\end{eqnarray}
The condition $\rho^\text{st}_e\ge0$ ensures that all other densities are non-negative as well. Therefore $p=p_{c}$ is the critical value for an absorbing transition with order parameter $\rho^\text{st}_e$: For $p\le p_{c}(\mu,r)$ the system reaches a stationary dynamically active phase in which $\rho_e^\text{st}$ is nonzero for all values of $r$, $\mu$ and $p$. In this regime a linear stability analysis indicates that this solution is linearly stable. On the other hand, for $p > p_{c}(\mu,r)$ the system reaches an absorbing or frozen state which turns out to be marginally stable. This analysis shows that the absorbing phase transition found in \cite{saeedian2019absorbing} for $r=0$ is robust under rewiring network dynamics.

\begin{figure}[]
\centering
\includegraphics[width=.9\textwidth]{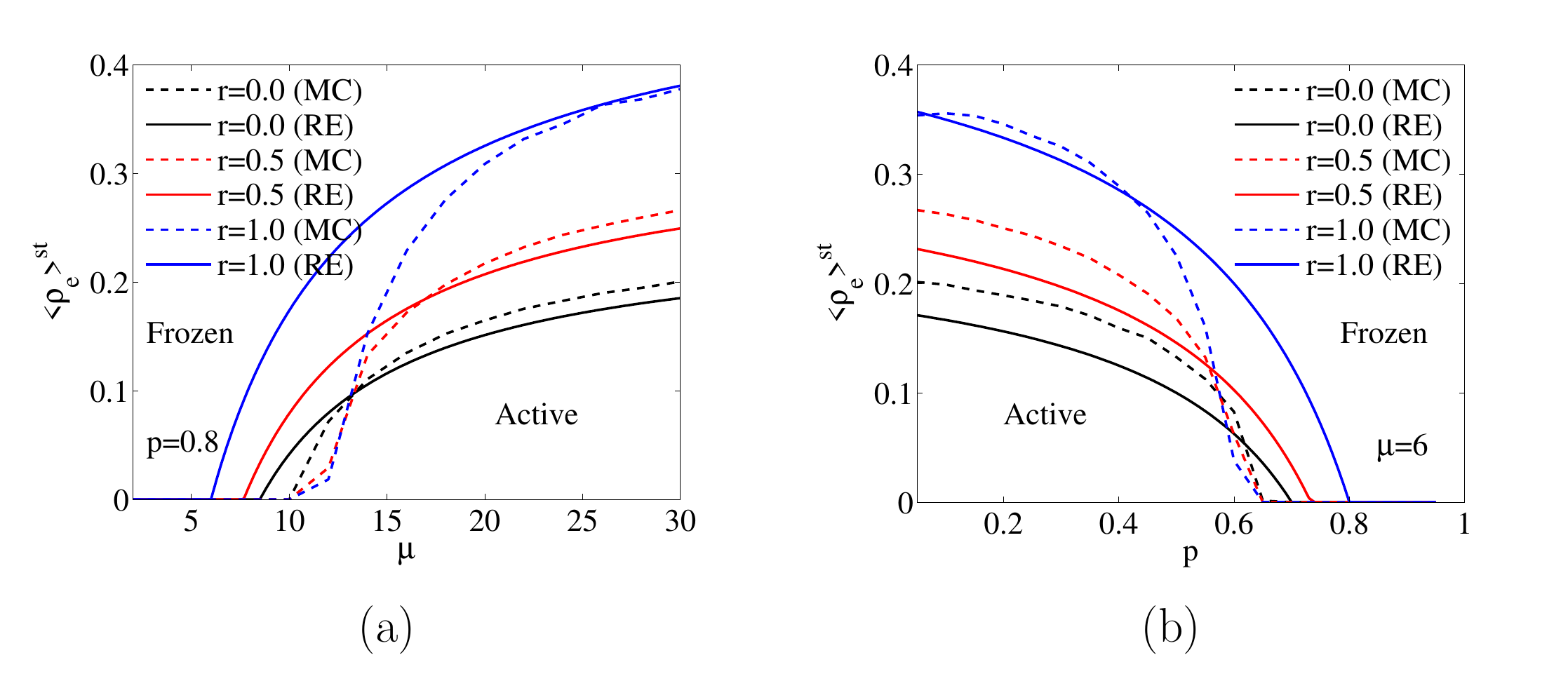}
\caption{Order parameter, $\langle\rho_{e}\rangle^{st}$, as a function of $\mu$ for fixed $p=0.8$, in panel (a), and as a function of $p$ for fixed $\mu=6$, in panel (b), in both cases for three values of the rewiring probability ($r=0, \ 0.5,\ 1$). The solid lines are the results of the rate equations while the dashed lines come from the Monte Carlo simulations on an Erd\H{o}s-R\'enyi network. Initial conditions for the densities $x_0$ of blue nodes and $\ell_0$ of attractive links are $(x_0,\ell_0)=(0.25,0.5)$. The system size is $N=500$ nodes, and averages are taken over 250 realizations.}
\label{cut_Ph_Di}
\end{figure}

\begin{figure}[]
 \centering
 \includegraphics[width=1\textwidth]{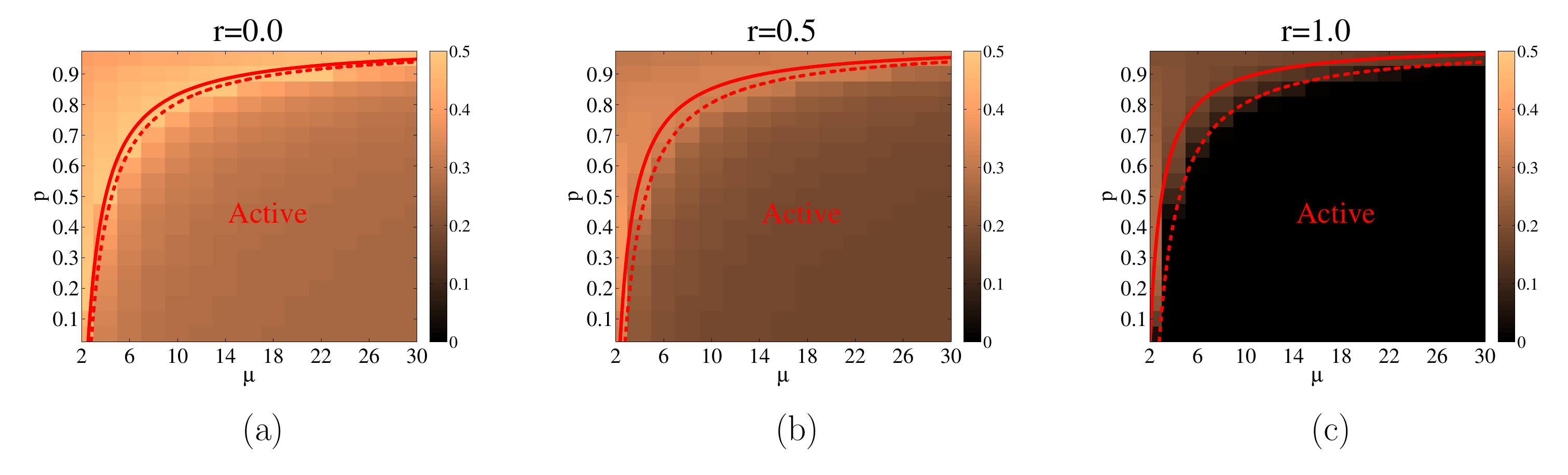}
\caption{Critical line for the active to absorbing transition in the $(p,\mu)$ plane for $r=0$, panel (a); $r=0.5$, panel (b); and $r=1$, panel (c). The red solid lines are the critical lines predicted by Eq.~\eqref{eq:pc} and the dashed lines are obtained from Monte Carlo simulations. The values of $\langle\rho_{f}\rangle^{st}$, obtained from the simulations are shown in a color scale. Initial conditions for the density $x_0$ of blue nodes and $\ell_0$ of attractive links are $(x_0,\ell_0)=(0.25,0.5)$. The system size is $N=500$ nodes, and averages are taken over 250 realizations.}
\label{rho_f_r}
\end{figure}

\begin{figure}[]
 \centering
 \includegraphics[width=.45\textwidth]{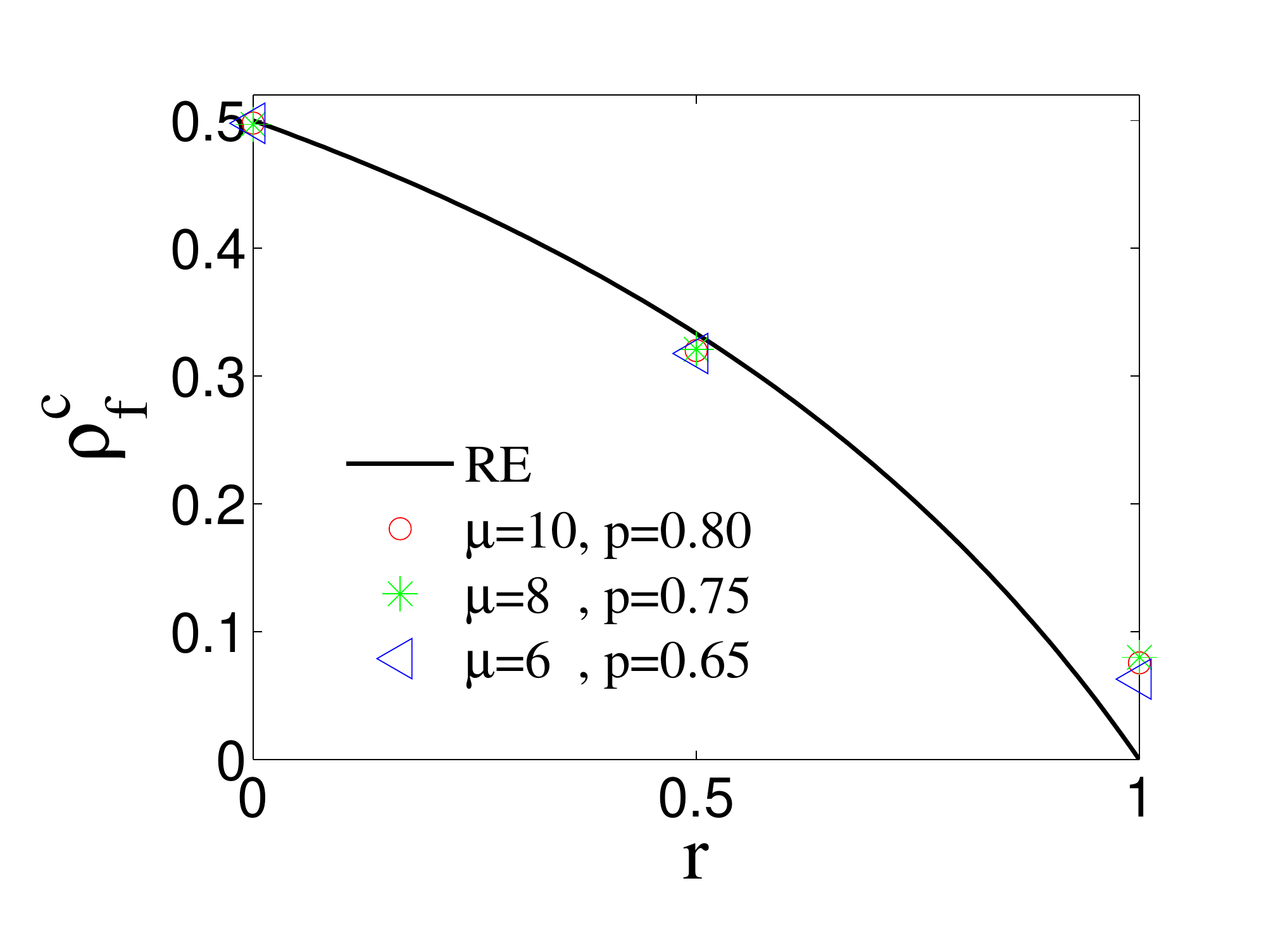}
\caption{The solid line is the value of $\rho_{f}^\text{st}$ at the critical line of the absorbing transition as a function of $r$, obtained from the analytical solution Eq.~\eqref{rhofc} coming from the rate equations. The markers for $r=0,0.5,1$ are obtained from Monte Carlo simulations in an Erd\H{o}s-R\'enyi network with $N=500$. The sets of $\mu$ and $p$ are associated with some different points on the critical line of absorbing transition (red dashed line in Fig.~\ref{rho_f_r})}.
\label{rho_f_r_c}
\end{figure}

In order to check our theoretical predictions of the active-to-absorbing transition we have conducted extensive Monte Carlo simulation of our coevolution model. The simulations have been carried out taking as initial condition an Erd\H{o}s-R\'enyi network with the desired average connectivity $\mu$. Initial conditions for the states of the nodes and states of the links are chosen randomly distributed with initial densities $x_0$ and $l_0$, respectively, of blue nodes and attractive links (see the \emph{Appendix}). Results of the simulations and the numerical integration of the rate equations are shown in Fig.~\ref{cut_Ph_Di} for three value of the rewiring probability ($r=0 ,\ 0.5,\ 1$) as a function of $\mu$ for $p=0.8$ fixed (left panel), and as a function of $p$ for $\mu=6$ (right panel). The Monte Carlo simulations confirm the predicted absorbing transition, but quantitative agreement for the order parameter $\rho_e^\text{st}$ in the active phase is better far from the transition point, that is, for large values of $\mu$ or small values of $p$. In addition, the transition point is shown to be rather independent of $r$. This fact is not captured by the approximate rate equations that, as mentioned earlier, assume a random regular network at all times. A comparison of the MC and RE results for the critical line in the $(p,\mu)$ plane for different values of $r$ is also shown in in Fig.~\ref{rho_f_r}, indicating a better overall agreement for small values of the probability $r$. This figure also shows in a color scale the values of $\rho_{f}^\text{st}$, the density of repulsive links connecting nodes in different states. This is a relevant quantity to monitor the effect of the network dynamics as measured by $r$: its value for all points in plane ($p,\mu$) decreases when increasing the rewiring parameter $r$ in such a way that in the active phase it becomes zero at $r=1$.
Moreover, the analytical solution of the rate equations, Eq.~\eqref{first_set}, predicts that the value of $\rho_{f}^\text{st}$ at the critical line $p=p_c$ is independent of $\mu$ and it is only a function of $r$. At the critical value of $p$, we find
\begin{eqnarray} 
q_c&=&(\mu-1)(1-p_{c}(\mu,r))=\frac{3-2r}{2-r},
\end{eqnarray}
independent of $\mu$. Replacing this value of $q_c$ in the expression for $\rho_{f}^\text{st}$ in Eq.~\eqref{first_set} we find the specially simple relation
\begin{eqnarray}\label{rhofc}
\rho_{f}^{c}(r)=\frac{1-r}{2-r}.
\label{rho_f_c}
\end{eqnarray}
This result, well confirmed by the MC simulations, is shown in Fig.~\ref{rho_f_r_c}, and it indicates a special behavior of the system for $r=1$. To be more precise, as we will discuss latter the mechanism of network fragmentation transition is rooted in the absence of pairs of type f at $r=1$ where $\rho_{f}^{c}\Big\rvert_{r=1}=0$.


\section*{The case r=1}
\begin{figure}[]
\centering
\includegraphics[width=.7\textwidth]{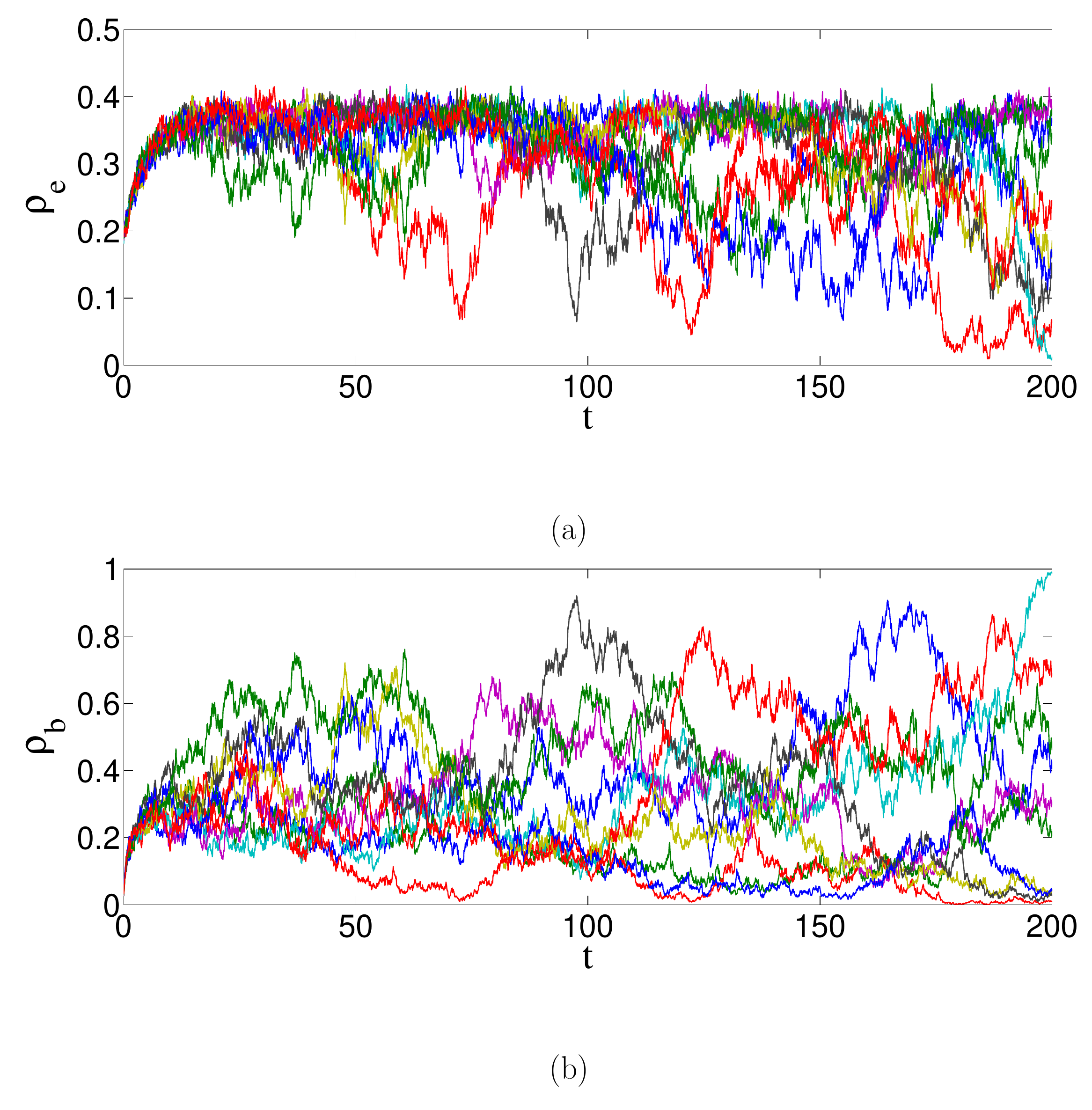}
\caption{Different realizations of the time evolution of $\rho_{e}$ ( panel (a)) and $\rho_{b}$ ( panel (b)) in the active phase for $r=1$ coming from numerical simulations on an Erd\H{o}s-R\'enyi network, with parameters $p=0.5$, $\mu=10$ and $N=500$ nodes. Initial conditions for the densities $x_0$ of blue nodes and $\ell_0$ of attractive links are $(x_0,\ell_0)=(0.25,0.5)$. Time is measured in MC steps.
}
\label{single_realization}
\end{figure}

\begin{figure}[]
 \centering
 \includegraphics[width=.4\textwidth]{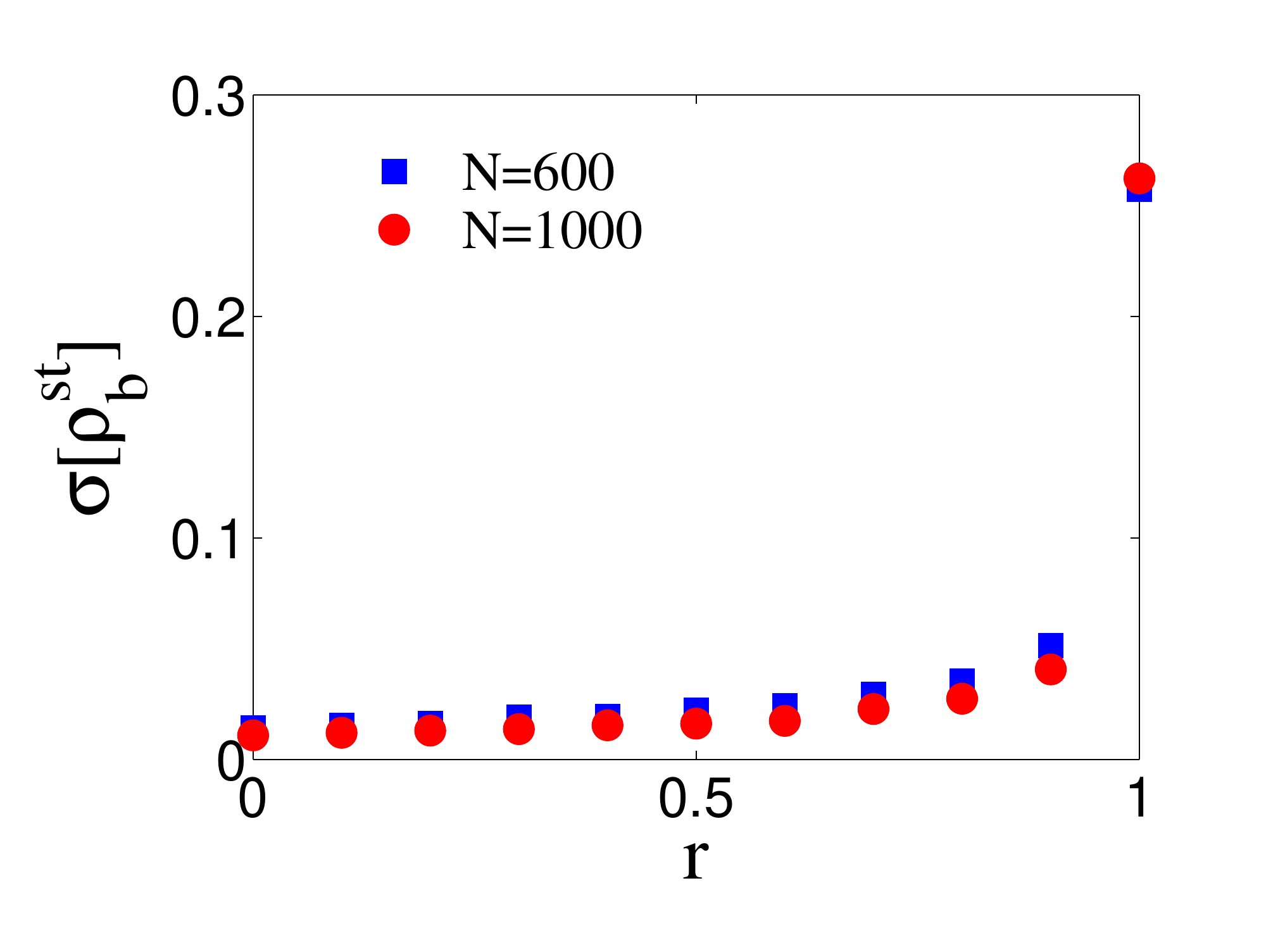}
\caption{Standard deviation of $\rho_{b}$ at the stationary active state as a function of the rewiring probability, $r$ for two systems sizes $N=600$ and $N=1000$. Parameters $p=0.5$, $\mu=10$.}
\label{sigma_b}
\end{figure}

When $r=1$ the unsatisfying pair \textit{e} only evolves by node state flipping or by link rewiring, but no link state flipping
is allowed. This turns out to be a singular situation in several aspects of our study. Firstly, we note that Eq.~\eqref{first_set}
implies that in the active phase the densities $\rho_a^\text{st}$, $\rho_c^\text{st}$ and $\rho_f^\text{st}$ are zero.
Therefore only one density of unsatisfying pairs ($\rho_e^\text{st}$) and two equivalent densities of satisfying pairs
($\rho_b^\text{st}$ and $\rho_d^\text{st}$) survive in this active phase. These non-vanishing dynamical variables exhibit very
large fluctuations as seen in 10 individual realizations of the process obtained by MC simulations
(Fig.~\ref{single_realization}). Starting from a fixed initial condition, fluctuations of $\rho_{b}$ and $\rho_{e}$
increase as time goes on. A quantitative measure of such fluctuations is given by the standard deviation of
$\rho_{b}$ at stationary state, $\sigma[\rho_{b}^\text{st}]$. We show in Fig.~\ref{sigma_b} the dependence of this quantity
on the rewiring probability, $r$, for two different system sizes. As illustrated, there is an abrupt increase of these fluctuations
at $r=1$, rather independent of system size.

Secondly, we note that in the absorbing phase it is always $\rho_a^\text{st}=\rho_e^\text{st}=\rho_c^\text{st}=0$. By continuity and on the critical line for $r=1$, $\rho_e^\text{st}=\rho_f^\text{st}=0$. Therefore in these conditions of criticality and $r=1$, the only non-vanishing densities are $\rho_b^\text{st}$ and $\rho_d^\text{st}$, satisfying the normalization condition $\rho_b^\text{st}+\rho_d^\text{st}=1$. Thus, there can be two types of solutions in these special conditions: (i) either $(\rho_b^\text{st},\rho_d^\text{st})=(0,1)$ or $(\rho_b^\text{st},\rho_d^\text{st})=(1,0)$, a situation with all nodes in the same state and only attractive links, or (ii) $\rho_b^\text{st}\ne0$ and $\rho_d^\text{st}\ne0$, which represents a fragmented network with disconnected groups, each of them with nodes in a given state connected by attractive links. This second solution, predicted by the rate equations in the thermodynamic limit, is a network fragmentation as a manifestation of criticality which occurs for $r=1$ as discussed in the last section on \textit{Topological transitions}.

\section*{Dynamics: Approach to the absorbing states and survival times}

\begin{figure}[]
\centering
 \includegraphics[width=0.7\textwidth]{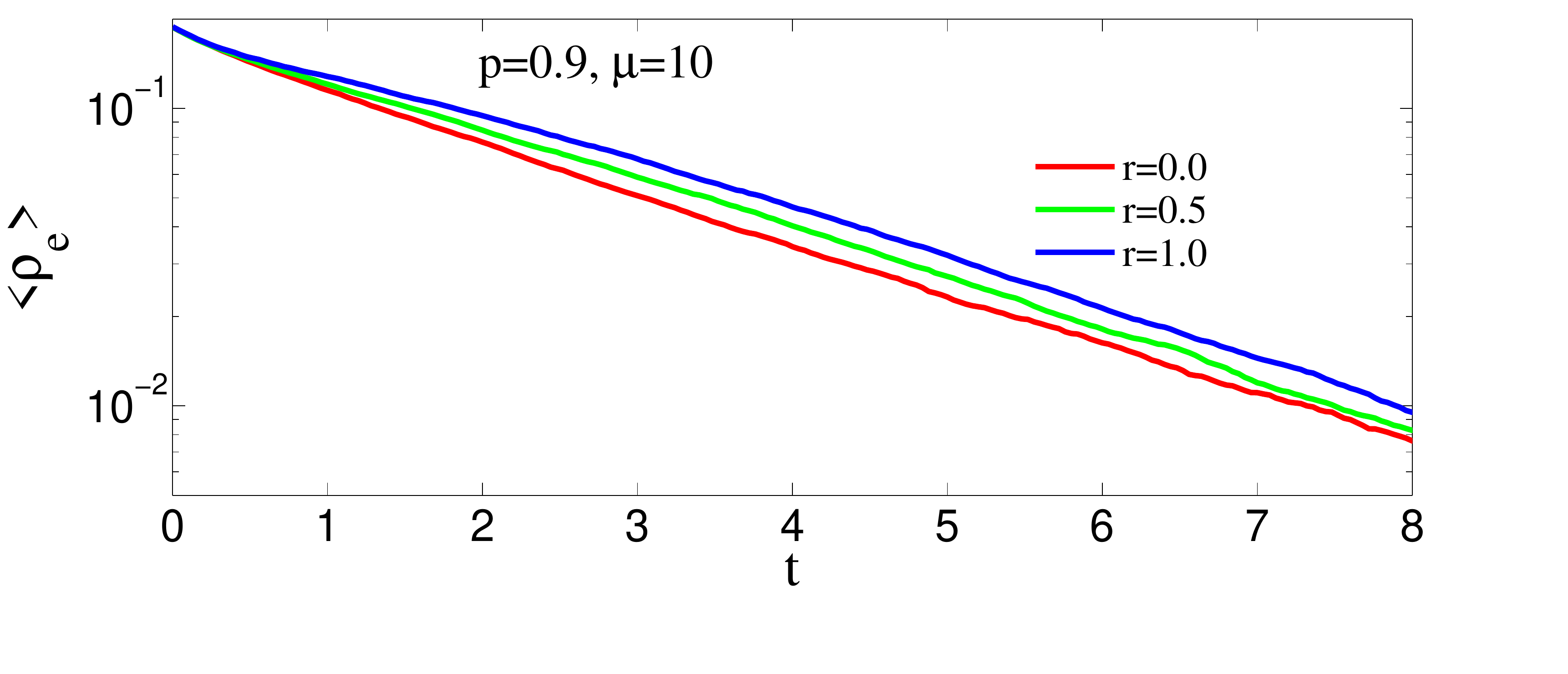}
\caption{Time evolution, obtained from MC simulations, of $\langle\rho_{e}\rangle$ in a log-linear scale for different values of $r$ as indicated. System parameters $p=0.9, \mu=10$ correspond to the absorbing phase. System size $N=500$. Initial conditions $x_0=0.25,l_0=0.5$. Averages taken over 240 MC realizations. Time measured in MC steps.}
\label{rho_e_t}
\end{figure}

\begin{figure}[]
 \centering
 \includegraphics[width=.9\textwidth]{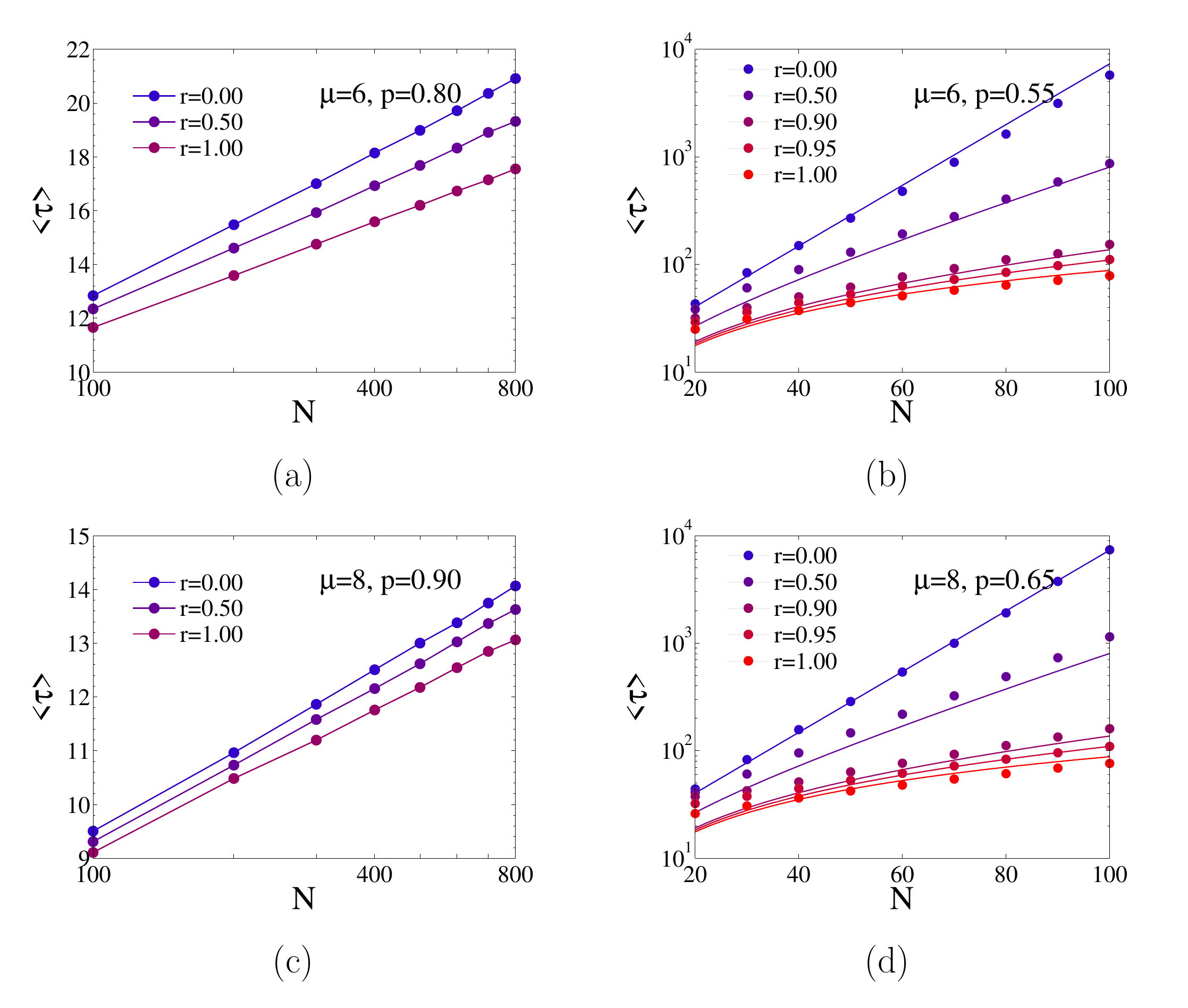}
\caption{Panels (a)\&(c): Absorbing phase. Mean value of the characteristic time, $\langle\tau\rangle$ to reach the absorbing state, as a function of system size for different values of rewiring probability, $r$, in a linear-log scale. Dots correspond to results of Monte Carlo simulations and the continuous lines are a logarithmic dependence fitting. Panels (b)\&(d): Active phase. Mean value of of the survival time, $\langle\tau\rangle$, as function of system size in a log-linear scale for different value of $r$. Dots correspond to the results of Monte Carlo simulations and the continuous lines are different fits: For $r=0$ an exponential fit, $\langle\tau\rangle\sim e^{\alpha N}$, for $r=1$ a linear fit, $\langle\tau\rangle\sim \beta N$, and for intermediate values of $r$ a phenomenological fit $\langle\tau\rangle= \alpha_1 (\alpha_2 N)^{r}e^{(1-r)\alpha_3 N}$ with $\alpha_1=11$, $\alpha_2=0.08$ and $\alpha_3=0.065$. In all panels the initial conditions for the densities of blue nodes and attractive links are: $x_0=0.5,l_0=0.5$. Averages are taken over 5,000 MC realizations. Time measured in MC steps.}
\label{tau}
\end{figure}

In the absorbing phase the system approaches an absorbing state from given initial conditions by an ordering process. This process is described by the time evolution of $\langle\rho_{e}\rangle$ as shown in Fig.~\ref{rho_e_t} for a set of parameters in the absorbing phase and three different values of the rewiring parameter $r$. We observe an exponential approach to the absorbing state, $\langle\rho_{e}\rangle \sim e^{-t/\tau_{0}}$. We also observe that $\tau_{0}$ is independent of $r$, so that the rewiring process does not modify the approach to the absorbing state. In a finite system of size $N$ the exponential approach lasts until $\langle\rho_{e}\rangle \sim 1/N$. This occurs at a characteristic time $\langle\tau\rangle \sim \log N$. This characteristic logarithmic dependence is shown in panels (a) and (c) of Fig.~\ref{tau} for three values of the rewiring probability and two sets of system parameters in the absorbing phase. Again, the logarithmic dependence is not modified by the rewiring process, although the value of $\langle \tau\rangle$ decreases with increasing $r$, indicating that the rewiring process accelerates the ordering process.

In the active phase of a finite system the dynamical state has a survival time because a finite-size fluctuation will eventually take
the system to an absorbing state. This survival time $\tau$ is a stochastic variable whose average value is shown in panels (b) and (d) of Fig.~\ref{tau} as a function of the system size for a set of parameters in the active phase and different values of $r$.
The curves for $r=0$ can be well fitted by an exponential dependence $\langle\tau\rangle\sim e^{\alpha N}$, while for $r=1$ we find a linear dependence $\langle\tau\rangle\sim \beta N$. For intermediate values of $r$ the scaling of $\langle\tau\rangle$ with $N$ interpolates between these two extreme dependencies. For given $r$, we conjuncture that the system size scaling of the survival time can be fitted by a phenomenological function as $\langle\tau\rangle= \alpha_1 (\alpha_2 N)^{r}e^{(1-r)\alpha_3 N}$. The chosen points (system parameters) of panels (b) and (d) in the plane ($p,\mu$) have the same distance from the critical line, and for both of them we find common fitting coefficients, $\alpha_1,\alpha_2,\alpha_3$. Therefore, in the active phase, as compared with the absorbing phase, the rewiring process has a much more significant effect in the dynamical processes changing the system size scaling. In addition a special behavior, the linear system size scaling of the survival time is found for $r=1$.

\section*{Topological transitions}

In a finite system, either in the absorbing phase or after the survival time in the active phase, the system reaches a fully satisfying configuration in which $\rho_a=\rho_c=\rho_e=0$. These configurations display a range of different non-trivial topological structures depending on the system parameters $(p,\mu,r)$. In this section we describe these configurations, the transitions among them and the relation with the active to absorbing transition discussed before.

\begin{figure}[]
\centering
 \includegraphics[width=.9\textwidth]{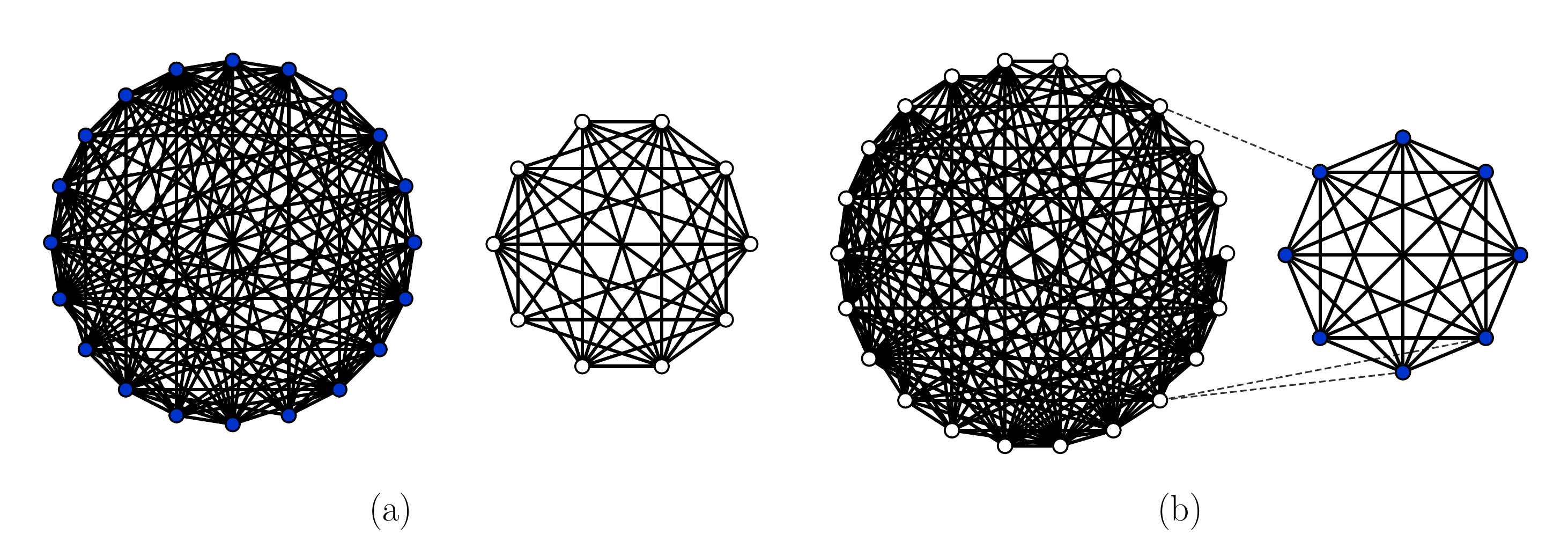}
\caption{(a) Example of a fragmented configuration obtained for $p=0.80$, $\mu=12$, $r=1$ and $N=30$. These parameters correspond to the critical line of the absorbing transition obtained from MC simulations as shown in Fig.~\ref{Frag_PS}. (b) Final two-group configuration obtained for the same parameters except for a slightly different rewiring probability, $r=0.98$. }
\label{snapshot_frag}
\end{figure}

\begin{figure}[]
\centering
 \includegraphics[width=.9\textwidth]{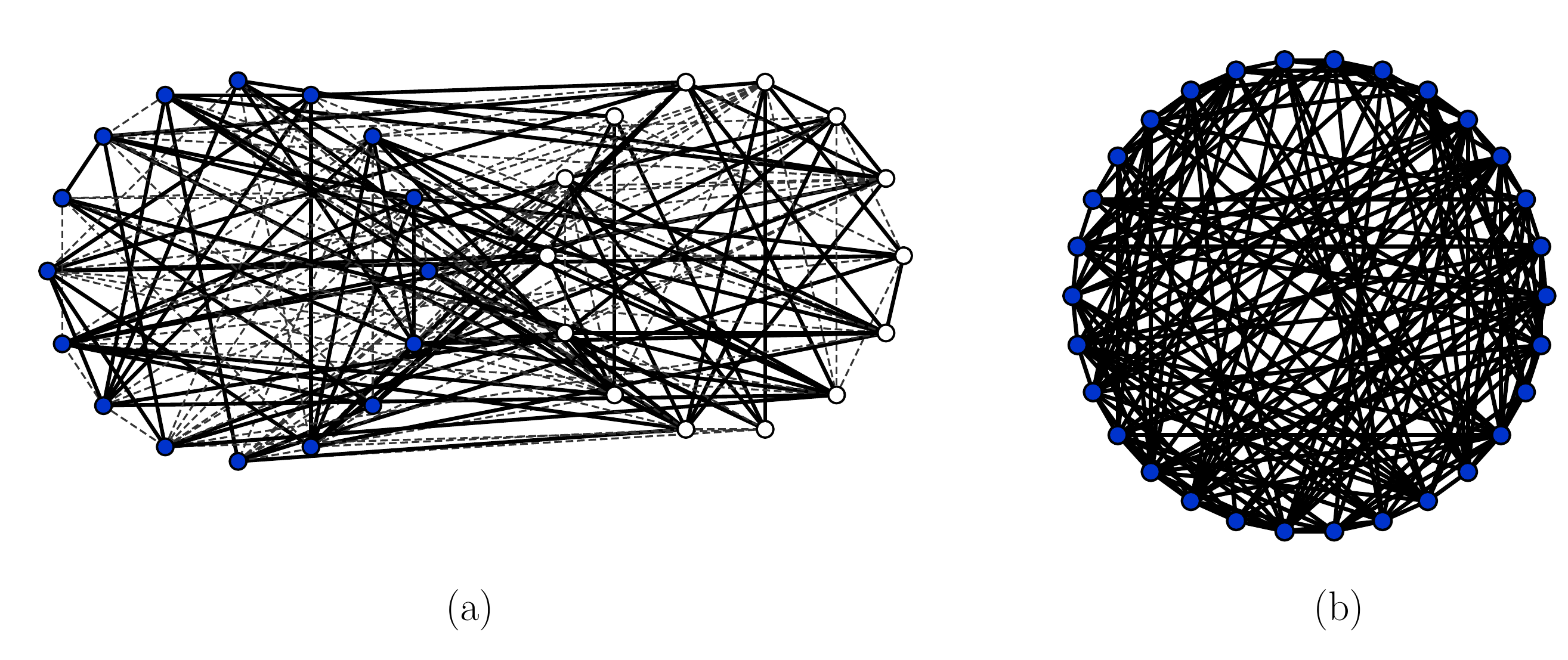}
\caption{(a) Snapshot of a dynamically active network in the active phase. (b) Final consensus configuration obtained after the survival time. Parameters values in the active phase $p=0.30$, $\mu=12$, $r=1$ and $N=30$.}
\label{snapshot_consus}
\end{figure}

\begin{figure}[]
\centering
 \includegraphics[width=1\textwidth]{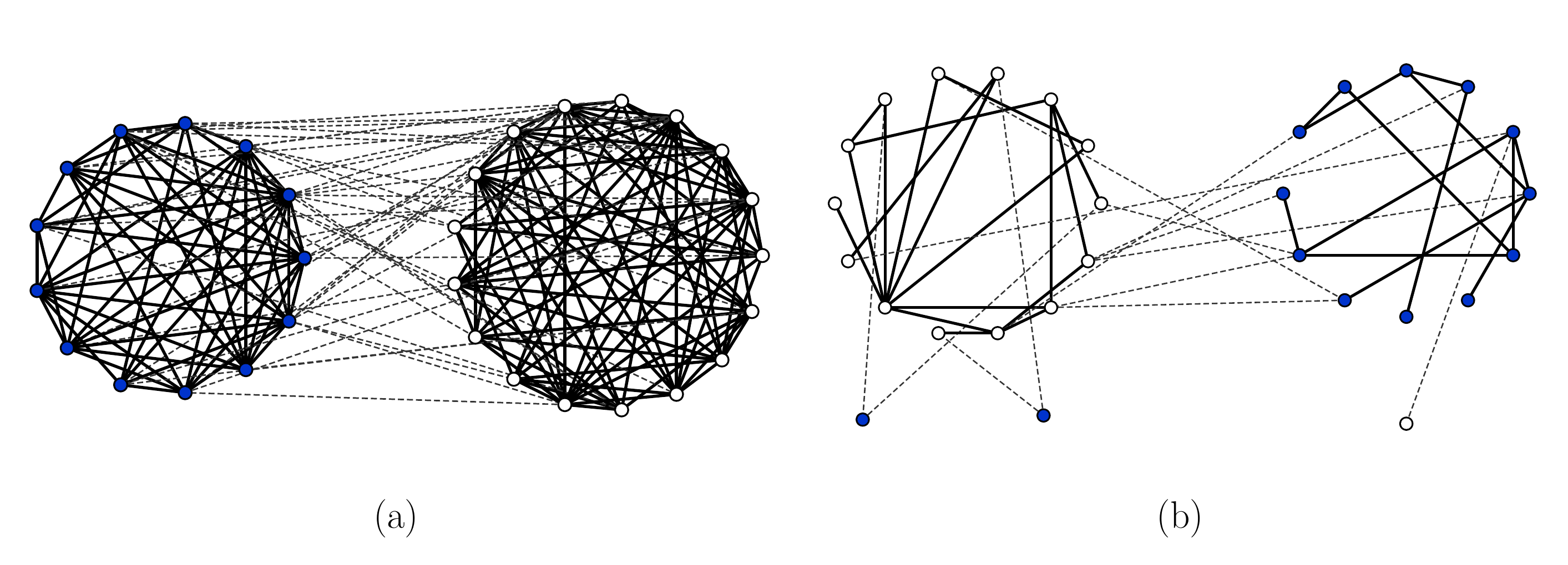}
\caption{(a) Two-group configuration obtained in the absorbing for $p=0.90$, $\mu=12$, $r=1$ and $N=30$. (b) Split configuration obtained in the absorbing phase for $p=0.30$, $\mu=3$, $r=1$ and $N=30$.}
\label{snapshot_two_split}
\end{figure}

The final, dynamically frozen, configurations can be classified according to the vanishing or non vanishing values of $\rho_b,\rho_d,\rho_f$. There are two main categories, fragmented and connected networks. The connected networks can be organized into three subclasses: consensus (CP), two-group (TP) and split configurations (SP). In the fragmented configurations it is $\rho_f=0$, $\rho_b\neq 0,\rho_d\neq 0$: therefore the nodes organize in two disjoint networks, one network containing all blue nodes and the ther all white nodes.  All links within the nodes of each independent network are attractive and there are no links between nodes belonging to different networks. An example of a fragmented configuration is shown in Fig.~\ref{snapshot_frag}(a). In the connected  consensus configurations, still $\rho_f=0$, but now either $\rho_b= 0$ or $\rho_d= 0$: all nodes are in the same state connected by attractive links as shown in Fiq.~\ref{snapshot_consus}(b). Both two-group and split configurations are characterized by $\rho_b\neq 0,\rho_d\neq 0,\rho_f\neq 0 $. In a two-group configuration, two groups of nodes, each group in a different state, are connected by some repulsive links, while the internal links of each group are attractive. An example is shown in Fiq.~\ref{snapshot_two_split}(a). The split configurations are similar to the two-group configurations except by the fact that some very small groups are attached to two large groups by some repulsive links as shown in Fiq.~\ref{snapshot_two_split}(b). 

\begin{figure}[]
\centering
 \includegraphics[width=1\textwidth]{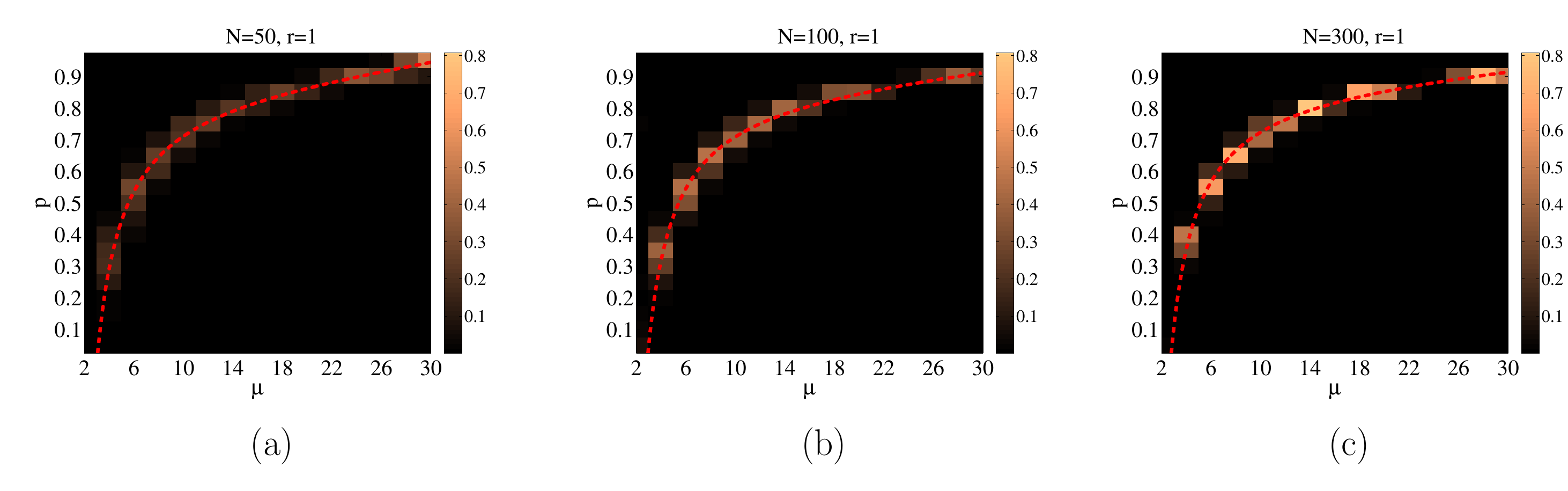}
\caption{Probability of finding a fragmented configuration (in color code ) as a function of $p$ and $\mu$ for $r=1$ and different system sizes: (a) $N=50$,(b)$N=100$,(c)$N=300$. The dashed red line indicates the critical line of the absorbing transition as obtained from MC simulations for the corresponding system size. This probability vanishes for all values of $p$ and $\mu$ except at the critical line. Initial conditions $x_0 = 0.5$, $l_0 = 0.5$. Averages taken over 500 MC realizations.}
\label{Frag_PS}
\end{figure}

\begin{figure}[]
\centering
 \includegraphics[width=1.1\textwidth]{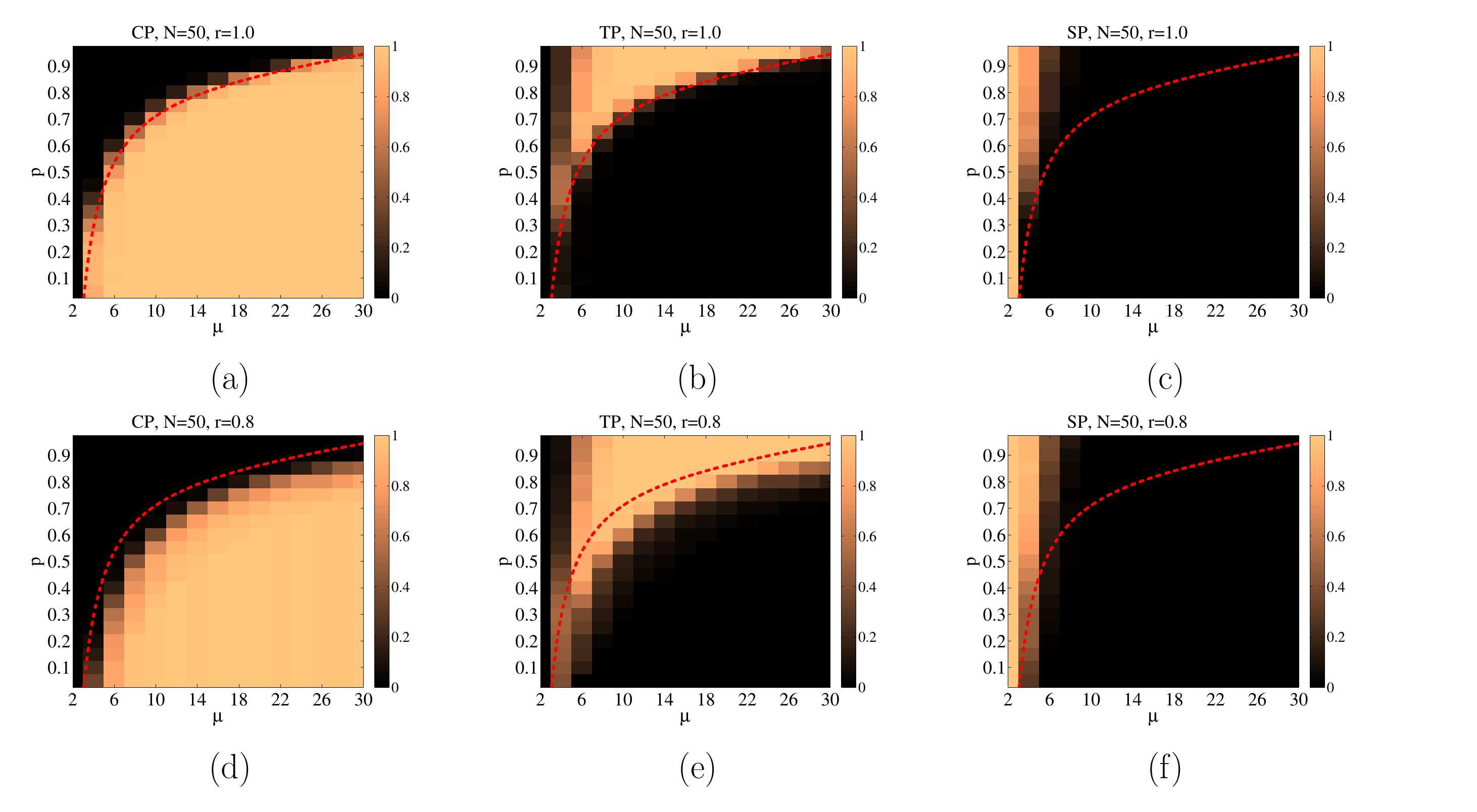}
\caption{The three top panels plot (in a color code) for $r=1$ the probabiliy of the system reaching asymptotically (a) a consensus configuration, (b) a two-group configuration, or (c) a split configuration. The three bottom panels plot the equivalent results for $r=0.8$. In all cases, the dashed red line indicates the critical line of the active-to-absorbing transition as obtained from MC simulations. System size $N=50$. Initial conditions $x_0 = 0.5$, $l_0 = 0.5$.
Averages taken over 500 MC realizations.}
\label{Connect_PS}
\end{figure}

We first consider which configurations are obtained for the special case $r=1$ for which we had interesting predictions from the rate equation analysis. The solution of the rate equations for $r=1$ in the active phase, and also at the critical line, predicts that $\rho_f=0$, which corresponds to either a consensus or a fragmented configuration. The Monte Carlo simulations conclude that the fragmented network solution is only found at the critical line of the absorbing transition, see an example in Fig.~\ref{snapshot_frag}(a). Further evidence of this result is shown in Fig.~\ref{Frag_PS}, where we plot, for different system sizes, the probability of occurrence of the fragmented structure in the ($p,\,\mu$) parameter plane, showing that this probability is only non-zero at the critical line and it increases with system size. An example of how this network fragmentation solution disappears as we move away from $r=1$ is shown in Fig.~\ref{snapshot_frag}(b). On the other hand, the consensus configuration is the one found in the active phase. An example of this configuration is shown in Fiq.~\ref{snapshot_consus}(b), together with a snapshot of the dynamically active network before it falls into the consensus absorbing state by a finite-size fluctuation. Evidence that, for $r=1$, the consensus solution is the only one found in the active phase, and that it also exists at the critical line, is given in Fig.~\ref{Connect_PS}(a)
that shows results of Monte Carlo simulations for the probability of occurrence of the consensus configuration in the ($p$,$\mu$)
parameter plane. As we move slightly away from $r=1$, e.g. Fig.~\ref{Connect_PS}(d) for $r=0.98$, the consensus configuration disappears for values of ($p$,$\mu$) in the active phase close to the critical line.

\begin{figure}[]
\centering
 \includegraphics[width=.8\textwidth]{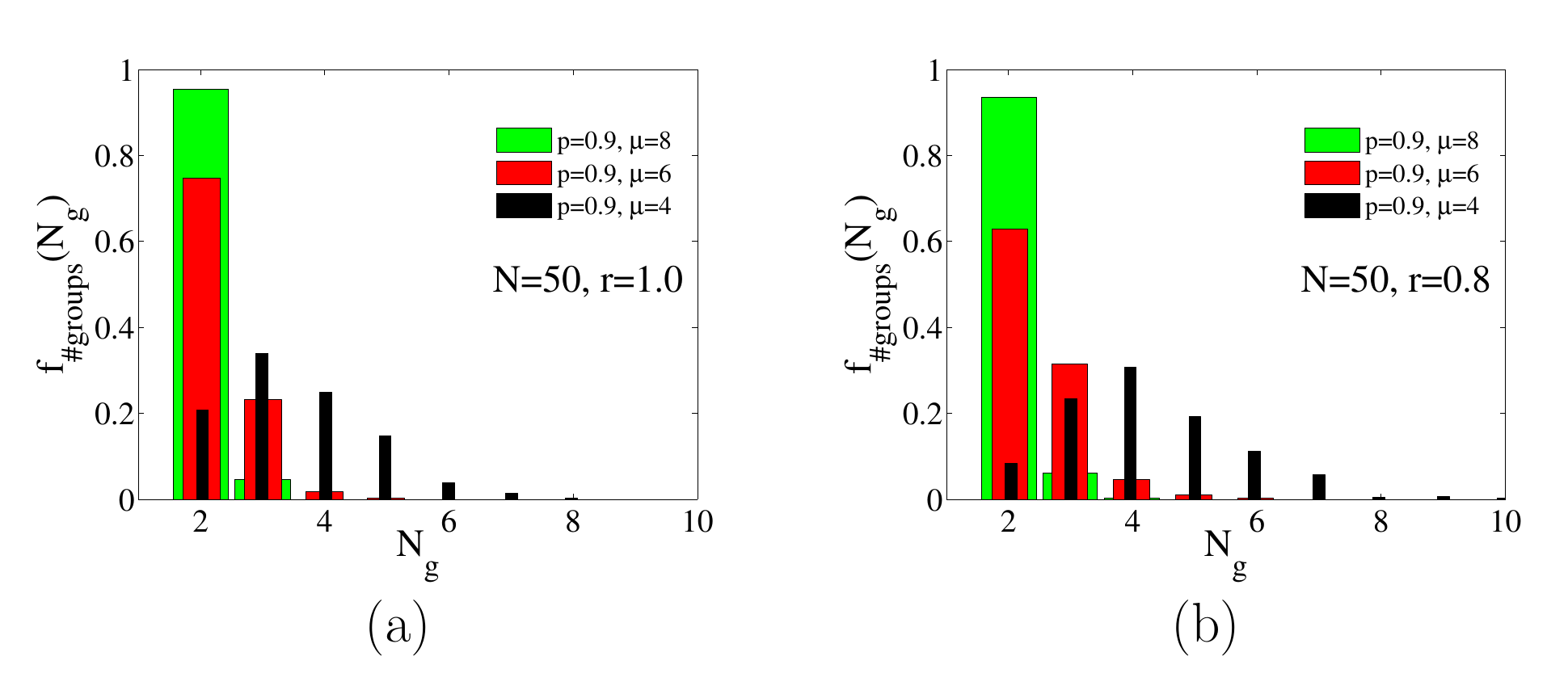}
\caption{Probability density function of the number of groups in the final network configuration for fixed $p$ but different values of $\mu$ as indicated. Panel (a) $r=1$ and panel (b) $r=0.8$. System size $N=50$. Initial conditions $x_0 = 0.5$, $l_0 = 0.5$.
Averages taken over 500 MC realizations.}
\label{TP_SP}
\end{figure}

In the absorbing phase for $r=1$, as well as in the absorbing and active phases when $r<1$, two-group and split configurations are found. The probability of occurrence of these solutions in the ($p$,$\mu$) parameter plane are shown for $r=1$ and $r=0.8$ in Fig.~\ref{Connect_PS}. We observe a transition between two-group and split configurations, named \emph{Finite-size topological transition} for $r=0$ in \cite{saeedian2019absorbing}. Our results indicate that this transition does not exist in the active phase for $r=1$, while for $r<1$, and for the chosen symmetric initial conditions, it occurs for a value of the network average degree $\mu_\text{split}$ which is essentially independent of $p$. In Fig.~\ref{TP_SP} we show the distribution of the number of groups for different parameter values. We use the precise definition of a group as a set of nodes in the same state, connected by attractive links among themselves and by repulsive links to the nodes of other groups. From this figure we estimate $\mu_\text{split}\sim 8$ which, as compared with values for $r=0$ \cite{saeedian2019absorbing}, indicates that $\mu_\text{split}$ decreases as a consequence of network dynamics ($r\neq 0$).



\section*{Summary and Discussion}

We have analyzed, by an analytical rate equation approach and by Monte Carlo simulations, a general dynamical model of interacting agents in a network in which nodes can be in two different states (blue or white) and links are either in an attractive or repulsive state. Unsatisfying pairs with repulsive links joining nodes in the same state, or with attractive links joining nodes in different states evolve to satisfying pairs by two different coevolving processes: i) Coupled dynamics of the state of the nodes and the state of the links; ii) network dynamics by random rewiring of an attractive link between two nodes in different states (pair \textit{e}) to connect to a node in the same state (pairs \textit{b} or \textit{d}). The rewiring parameter $0\leq r \leq1$ is such that there is no dynamics of the topology of the network for $r=0$, while for $r=1$ pair \textit{e} only evolves by updating the state of the nodes or by link rewiring, but there is no change of the state of the link from attractive to repulsive. We find a phase transition, predicted by the rate equations, between an active phase with persistent dynamics and an absorbing phase in which the system has reached an absorbing configuration as steady state. This phase transition is already found for $r=0$ \cite{saeedian2019absorbing}, but its existence is robust against the introduction of the network dynamics: The origin of the transition and the critical line do not significantly depend on the rewiring parameter $r$. However, the network dynamics in the limiting case $r=1$ produces very significant physical changes at criticality and in the active phase. As predicted by the rate equations, we find network fragmentation on the critical line, and only on this line, so that fragmentation is a manifestation of criticality. In the active phase only links in the attractive state survive and the fluctuations of the density of attractive links connecting nodes in the same state diverges as $r\longrightarrow 1$ indicating the singularity of the case $r=1$. Finite-size fluctuations in the active phase take the system, after a survival time, to a consensus frozen configuration with all nodes in the same state and connected by attractive links, while for $r\neq 1$ there are other possible configurations with a finite-size topological transition between two-group and split configurations. In addition, the survival time scales linearly with system size $N$, while in the absence of link rewiring it scales exponentially with $N$.

We have found four possible final frozen configurations of the system: fragmented, consensus,  two-group and split. When our model is understood as an opinion formation model, these configurations, except the consensus one, can be interpreted as a manifestation of social polarization. Our results give support to the claim that negative interactions are a major cause
of social polarization. Indeed, our solution for $r=1$ in the active phase is such that repulsive
links do not survive and the consequence is that the final state is that of consensus.

\section*{Appendix}
\subsection*{Rate equations for the coupled evolution of node and link states and rewiring process}
The dynamical behavior of the different densities of pairs as a function of time can be obtained as a set of so-called rate equations whose derivation we now sketch. The basic ingredient is the evolution rules sketched in Fig.~\ref{update_rule_Co} as considered on a network in which each node has exactly $\mu$ links at all times. This assumption is a clear approximation because in the rewiring process the degrees of the nodes change over time. With this provision in mind, the rate equations are written as 
\begin{eqnarray}\label{eq:rate}
\frac{d\rho_{a}}{dt}&=&-\rho_{a}+(1-p)(\mu-1)\Big(-\frac{\rho_{a}^{2}}{\chi}+\frac{1}{2}\frac{\rho_{c}\rho_{f}}{\phi}+\frac{1}{2}\frac{\rho_{e}\rho_{f}}{2\phi}-\frac{\rho_{e}\rho_{a}}{2\chi}\Big), \cr
\frac{d\rho_{b}}{dt}&=& p\rho_{a}+\frac{1-p+rp}{2}\rho_{e}+(1-p)(\mu-1)\Big(-\frac{\rho_{a}\rho_{b}}{\chi}+\frac{1}{2}\frac{\rho_{c}\rho_{e}}{\phi}+\frac{1}{2}\frac{\rho_{e}^{2}}{2\phi}-\frac{\rho_{e}\rho_{b}}{2\chi}\Big), \cr
\frac{d\rho_{c}}{dt}&=&-\rho_{c}+(1-p)(\mu-1)\Big(+\frac{1}{2}\frac{\rho_{a}\rho_{f}}{\chi}-\frac{\rho_{c}^{2}}{\phi}+\frac{1}{2}\frac{\rho_{e}\rho_{f}}{2\chi}-\frac{\rho_{e}\rho_{c}}{2\phi}\Big), \cr
\frac{d\rho_{d}}{dt}&=& p\rho_{c}+\frac{1-p+rp}{2}\rho_{e}+(1-p)(\mu-1)\Big(+\frac{1}{2}\frac{\rho_{a}\rho_{e}}{\chi}-\frac{\rho_{c}\rho_{d}}{\phi}+\frac{1}{2}\frac{\rho_{e}^{2}}{2\chi}-\frac{\rho_{e}\rho_{d}}{2\phi}\Big), \cr
\frac{d\rho_{e}}{dt}&=& -\rho_{e}+(1-p)(\mu-1)\Big(-\frac{1}{2}\frac{\rho_{a}\rho_{e}}{\chi}+\frac{\rho_{a}\rho_{b}}{\chi}+\frac{\rho_{c}\rho_{d}}{\phi}-\frac{1}{2}\frac{\rho_{c}\rho_{e}}{\phi}+\frac{\rho_{e}\rho_{d}}{2\phi} \cr
&&-\frac{1}{2}\frac{\rho_{e}^{2}}{2\chi}-\frac{1}{2}\frac{\rho_{e}^{2}}{2\phi}+\frac{\rho_{e}\rho_{b}}{2\chi}\Big), \cr
\frac{d\rho_{f}}{dt}&=&(1-r)p\rho_{e}+(1-p)(\rho_{a}+\rho_{c})+(1-p)(\mu-1)\Big(-\frac{1}{2}\frac{\rho_{a}\rho_{f}}{\chi}+\frac{\rho_{a}^{2}}{\chi}+\frac{\rho_{c}^{2}}{\phi} \cr
&&-\frac{1}{2}\frac{\rho_{c}\rho_{f}}{\phi}+\frac{\rho_{e}\rho_{c}}{2\phi}-\frac{1}{2}\frac{\rho_{e}\rho_{f}}{2\chi}-\frac{1}{2}\frac{\rho_{e}\rho_{f}}{2\phi}+\frac{\rho_{e}\rho_{a}}{2\chi} \Big), \cr
\chi&=&\rho_{a}+\rho_{b}+\rho_{e}+\rho_{f}, \cr
\phi&=&\rho_{c}+\rho_{d}+\rho_{e}+\rho_{f}.
\end{eqnarray}
The only pairs chosen by the update rule are $a$, $c$ and $e$. However, the update of one of these pairs will also change the number of pairs $b$, $d$ and $f$, as in the process of node update, the states of links connected to the updated node will also changed. The nonlinear terms in the rate equation account for this mechanism. In the following we explain the derivation of the rate equations considering the linear and nonlinear terms separately.

The linear terms are obtained by the variation of densities due to the direct update of the state of nodes and links in a time step as
\begin{eqnarray}
\frac{d\rho_i}{dt}=+\sum_{j}\frac{d\rho_i}{dt}\Big\rvert_{i\rightarrow j}+[nonlinear \ terms], \label{deriv_i} \\
\frac{d\rho_j}{dt}=-\sum_{i}\frac{d\rho_i}{dt}\Big\rvert_{i\rightarrow j}+[nonlinear \ terms], \label{deriv_j}
\end{eqnarray}
where $\frac{d\rho_i}{dt}\Big\rvert_{i\rightarrow j}$ is the direct effect of the update from pair $i\in \{a,c,e\}$ to the $j\in \{b,d,f\}$.

As an example, we now we derive in detail the linear terms of the equation for $\rho_{e}$: In a given update step, with probability $\rho_{e}$ a pair \textit{e} is randomly chosen and updated producing a density changes of $\Delta\rho_e=-\frac{1}{N\mu/2}$. According to the local adaptation rule in Fig.~\ref{update_rule_Co}, with probability $(1-r)p$, pair \textit{e} turns into the pair \textit{f}. With probability $\frac{1-p}{2}$, the pair \textit{e} becomes \textit{b} and with probability $\frac{1-p}{2}$, it becomes \textit{d}. In addition, according to the rewiring update rule in Fig.~\ref{update_rule_Co}, with probability $\frac{rp}{2}$, the pair \textit{e} becomes \textit{b} and with probability $\frac{rp}{2}$, it becomes \textit{d}. The time interval (measured in Monte Carlo steps) in any update step is given by $\Delta t=\frac{1}{N\mu/2}$. Therefore, the variation of $\rho_{e}$ due to the direct effect of update of the pair \textit{e} is
\begin{eqnarray}
\frac{d\rho_e}{dt}&=&\frac{d\rho_e}{dt}\Big\rvert_{e\rightarrow f}+\Bigg(\frac{d\rho_e}{dt}\Big\rvert_{e\rightarrow b}^{local}+\frac{d\rho_e}{dt}\Big\rvert_{e\rightarrow b}^{non-local}\Bigg)+\Bigg(\frac{d\rho_e}{dt}\Big\rvert_{e\rightarrow d}^{local}+\frac{d\rho_e}{dt}\Big\rvert_{e\rightarrow d}^{non-local}\Bigg) \cr
&=&(1-r)p\frac{\Delta\rho_e}{\Delta t}\rho_e+\frac{1-p}{2}\frac{\Delta\rho_e}{\Delta t}\rho_e+\frac{rp}{2}\frac{\Delta\rho_e}{\Delta t}\rho_e+\frac{1-p}{2}\frac{\Delta\rho_e}{\Delta t}\rho_e+\frac{rp}{2}\frac{\Delta\rho_e}{\Delta t}\rho_e \cr
&=&-\rho_{e}.
\label{deriv_exa}
\end{eqnarray}

Non-linear terms are an indirect effect of the node update. Blue nodes can be an end to any of the links \textit{c}, \textit{d}, \textit{e} and \textit{f} while white nodes can be an end to any of the links \textit{a}, \textit{b}, \textit{e} and \textit{f}. Thus, node update will change the state of the connected links to the updated node so that
\begin{eqnarray}
\frac{d\rho_{v}}{dt}&=&[linear \ terms]+\sum_{i,j}\bigg(\frac{d\rho_v}{dt}\Big\rvert_{i\to j}^{v\to w}-\frac{d\rho_w}{dt}\Big\rvert_{i\to j}^{w\to v}\bigg), \cr
\frac{d\rho_{w}}{dt}&=&[linear \ terms]+\sum_{i,j}\bigg(\frac{d\rho_w}{dt}\Big\rvert_{i\to j}^{w\to v}-\frac{d\rho_v}{dt}\Big\rvert_{i\to j}^{v\to w}\bigg).
\end{eqnarray}
where $\frac{d\rho_v}{dt}\Big\rvert_{i\to j}^{v\to w}$ and its negative value are the change in density of the pairs $v\in \{a,b,c,d,e,f\}$ and $w\in \{a,b,c,d,e,f\}$, respectively, due to the node update of the pair $i$ to $j$. Note that some terms like $\frac{d\rho_e}{dt}\Big\rvert_{a\to f}^{e\to b}$ are zero.

\begin{figure}\centering
\includegraphics[width=.6\linewidth]{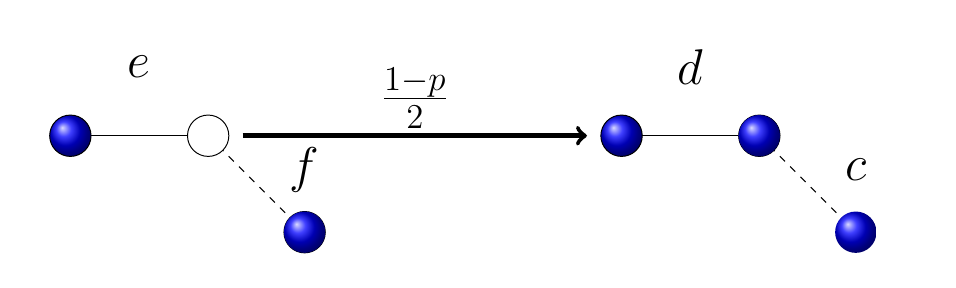}
\caption{The pair \textit{f} becomes \textit{c} when a node update leading from pair \textit{e} to pair \textit{d} takes place.}
\label{non-linear_term}
\end{figure}

As an example of a specific case of how to obtain the nonlinear terms, we derive the non linear term $-(1-p)(\mu-1)\frac{1}{2}\frac{\rho_{e}\rho_{f}}{2\chi}$ in the equation for $\rho_{f}$. This term is the consequence of a node update from the pair connection \textit{e} to \textit{d}. In a Monte Carlo step, with probability $\rho_{e}$ a pair \textit{e} is randomly picked and with probability $\frac{1-p}{2}$ a node update to became \textit{d} takes place. Now, let us examine the change of the rate of $\rho_{f}$ under the update of \textit{e} to \textit{d} as presented in Fig.~\ref{non-linear_term}. The normalized number of whole pair connections attached to the one side of a pair \textit{e} is $\frac{\mu-1}{N\mu/2}$ and from this portion, the fraction of the pair \textit{f} that is attached to the link \textit{e} is $\frac{\rho_{f}}{\chi}\frac{\mu-1}{N\mu/2}$. In addition, due to the asymmetry of pairs \textit{e} and \textit{f}, the update from any side of these pairs would result in a different pair. For instance, if the white opinion in the pair \textit{f} is flipped it converts to \textit{c}, while if the blue opinion is flipped, it turns to \textit{a}. Thus, when we deal with pairs \textit{e} and \textit{f}, the contribution of nonlinear terms appear with probability $\frac{1}{2}$. Thus, the global change in the density of pairs \textit{f} due to the node update from \textit{e} to \textit{d} is given by $\Delta\rho_f=-\frac{\rho_{f}}{2\chi}\frac{\mu-1}{N\mu/2}$.
\begin{eqnarray}
\frac{d\rho_f}{dt}\Big\rvert_{e\to d}^{f\to c}=\frac{1-p}{2}\frac{\Delta\rho_f}{\Delta t}\rho_e =-\frac{1-p}{2}(\mu-1)\frac{\rho_{e}\rho_{f}}{2\chi}
\end{eqnarray}

\subsection*{Initial condition}
In the numerical integration of Eqs.\eqref{eq:rate} it is important to ensure that the initial conditions satisfy the relations 
\begin{eqnarray}
\rho_{a}+\rho_{b}&=&\frac{n\mu_w}{N\mu}=x\frac{\mu_w}{\mu} \nonumber \\
\rho_{c}+\rho_{d}&=&\frac{(N-n)\mu_b}{N\mu}=(1-x)\frac{\mu_b}{\mu} \label{rho_rel}\nonumber \\
\rho_{e}+\rho_{f}&=&\frac{N\mu-(N-n)\mu_b-n\mu_w}{N\mu}=(1-x)\frac{\mu_b}{\mu}-x\frac{\mu_w}{\mu}
\label{mu_w_b}
\end{eqnarray}
where $\mu_b$ and $\mu_w$ stand for the mean degree of the blue and white nodes, respectively. Note that, $\mu_b$ and $\mu_w$ are functions of time but the total number of links in the network, $ \frac{N\mu}{2}$, is a conserved quantity. In the numerical integration of the rate equations and the MC simulation we choose regular random and Erd\H{o}s-R\'enyi networks as initial configurations, respectively. In both cases, requirements on initial conditions are satisfied taking
\begin{eqnarray}
\rho_a(0)&=&x_0^2(1-\ell_0)\nonumber\\
\rho_b(0)&=&x_0^2\ell_0\nonumber\\
\rho_c(0)&=&(1-x_0)^2(1-\ell_0)\nonumber\\
\rho_d(0)&=&(1-x_0)^2\ell_0 \nonumber\\
\rho_e(0)&=&2x_0(1-x_0)\ell_0 \nonumber\\
\rho_f(0)&=&2x_0(1-x_0)(1-\ell_0)
\end{eqnarray}
where $\ell_0$ is the initial fraction of attractive links and $x_0$ the initial fraction of blue nodes. 

Finally, in the numerical integration of the rate equations, we employed the predictor-corrector method. 

\section*{References}

\bibliographystyle{unsrt}
\bibliography{Draft}

\section*{Acknowledgements}
Financial support has been received from the Agencia Estatal de Investigacion (AEI, MCI, Spain) and Fondo Europeo de Desarrollo Regional (FEDER, UE), under Project PACSS (RTI2018-093732-B-C21/C22) and the Maria de Maeztu Program for units of Excellence in R\&D (MDM-2017-0711).

\end{document}